\documentclass[prb,aps,eqsecnum,showpacs,twocolumn]{revtex4}

\usepackage[dvips]{graphics}
\usepackage{epsfig}
\usepackage{graphicx}
\usepackage{amssymb}
\usepackage{epstopdf}
\usepackage{bm}
\DeclareGraphicsExtensions{.bmp,.eps,.gif,.png}

\begin{document}


\title{Ballistic transport in ferromagnet--superconductor--ferromagnet
trilayers with arbitrary orientation of magnetizations}

\author{Milo\v{s} Bo\v{z}ovi\'{c} and Zoran Radovi\'c}

\affiliation{Department of Physics, University of Belgrade,
P.O. Box 368, 11001 Belgrade, Serbia and Montenegro}

\begin{abstract}
Transport phenomena in clean ferromagnet--superconductor--ferromagnet
(FSF) trilayers are studied theoretically for a general case of arbitrary
orientation of in-plane magnetizations and interface transparencies.
Generalized expressions for scattering probabilities are derived
and the differential conductance is computed using solutions of the Bogoliubov--de Gennes equation. We focus on size and coherence
effects that characterize ballistic transport, in particular on the subgap transmission and geometrical oscillations of
the conductance. We find a monotonic dependence of
conductance spectra and magnetoresistance on the angle of
misorientation of magnetizations as their alignment is changed from
parallel to antiparallel. Spin-triplet pair correlations in FSF
heterostructures induced by non-collinearity of magnetizations are
investigated by solving the Gor'kov equations in the clean limit.
Unlike diffusive FSF junctions, where the triplet correlations have
a long-range monotonic decay, we show that in clean
ferromagnet-superconductor hybrids both singlet and triplet
pair correlations induced in the F layers are oscillating and
power-law decaying with the distance from the S-F interfaces.
\end{abstract}


\maketitle

\section{Introduction}

The interplay between ferromagnetism and superconductivity in thin
film systems is a phenomenon that attracts considerable interest of
researchers for some time already.\cite{Buzdin_rev} Apart from
potential device applications, ferromagnet--superconductor (F--S)
hybrid structures can be major tools in the quest for novel
superconducting phenomena.\cite{Pokrovsky_rev,BVE_rev} Variety of
interesting theoretical predictions, such as the existence of
$\pi$-state superconductivity in F--S multilayer
systems,\cite{Bulaevski,FFLO,Buzdin82} and characteristic
oscillations in the superconducting transition temperature as a
function of thickness of ferromagnetic
layers,\cite{Radovic91,Buzdin90,Demler,Tagirov_C,FCG,Bagrets} have
been already confirmed
experimentally,\cite{Kontos,Ryazanov,Jiang,Lazar,Garifullin,Obi}
while the existence of the spin valve
effect\cite{Tagirov_PRL,Baladie} was more difficult for confirmation
than expected.\cite{Gu,PRL2005} Other peculiar phenomena that have
been predicted to exist in F--S hybrids with inhomogeneous
magnetization -- the occurrence of a long-range triplet
pairing,\cite{F1,BVE_PRL,F2,KK,VBE} or the so-called inverse
proximity effect\cite{BVE_PRB} -- still wait for experimental
realization. Hence, a proper understanding of these effects is a
prerequisite for setting up appropriate experimental conditions.

Long-range triplet pairing, in particular, was first predicted as a
consequence of proximity of an inhomogeneous ferromagnet to a
superconductor.\cite{F1,BVE_PRL,F2,KK} However, it was shown that
spin-triplet pair correlations may also arise in layered structures consisting
of superconductors and homogeneous ferromagnets with differently
oriented magnetizations.\cite{VBE,Kupriyanov} The simplest structure
of this kind is an FSF system with homogeneous but non-collinear
magnetizations of the ferromagnets. In the parallel (P) or the
antiparallel (AP) alignment such correlations are absent. It was found that in
diffusive junctions triplet correlations have a
long-range monotonically decaying component in the ferromagnetic
layers.\cite{VBE} Very recently, effects of these correlations seem to have been observed in half-metallic ferromagnets.\cite{Keizer,Penya,Eschrig03}

In this paper we study a variety of effects that occur due to the
interplay of ferromagnetism and superconductivity in clean FSF
double junctions. It has been already established for the case of
collinear magnetizations that the most significant consequences of
quantum interference in such structures are the subgap transmission
and the periodic vanishing of the Andreev reflection.\cite{BozovicB}
The annulment of the Andreev reflection occurs at the energies of
geometrical resonances, which correspond to the maxima in the
density of states. Here, we investigate these phenomena in FSF
hybrids with an arbitrary angle between magnetizations. In
particular, we address the issue of triplet correlations, and
describe its influence on transport properties.

We use solutions of the Bogoliubov--de Gennes equation in the
scattering formulation (Section II) to obtain the probabilities of
processes that charge carriers undergo and calculate the
conductance spectra for arbitrary orientation of magnetizations.
Section III illustrates some typical cases. In particular, the
conductance spectra of a clean FSF hybrid with a thin
superconducting layer (i.e., such that the layer thickness $d_s$ is
less or of the order of the superconducting coherence length
$\xi_0=\hbar v_{\rm F}/\pi \Delta_0$) are dominated by subgap
transport since majority of the charge carriers is transferred
directly from one electrode to another without interaction with
superconducting condensate.\cite{BozovicB,Yamashita67} On the other
hand, situation in the junctions with a thick superconducting layer
($d_s\gg\xi_0$) is quite the opposite: majority of the carriers form
Cooper pairs, thereby increasing the subgap conductance; also, the
conductance spectra show pronounced oscillatory behavior above the
gap due to interference-generated geometric resonances inside the
superconductor. However, we show that no extraordinary effects arise
when the relative orientation of magnetizations is between parallel
and antiparallel -- the spectra for intermediary angles simply fall
in between those for P and AP alignment. The obtained results allow
us to infer some general conclusions about the nature of
ballistic transport properties of FSF hybrids, which we illustrate
on the example of dependence of magnetoresistance on voltage and
S-layer thickness (Section IV). In Section V we address the problem
of spin-triplet pair correlations when magnetizations are non-collinear.
By solving the Gor'kov equations in the clean limit, we show that
both triplet and singlet superconducting correlations induced in ferromagnets are
oscillatory functions of distance from S-F interfaces, with a power law decay. This result is consistent with obtained
monotonicity of transport properties as the angle of relative
orientation of magnetizations is varied.

\section{The model}

\begin{figure}[h]
\begin{center}
    \includegraphics[width=7cm]{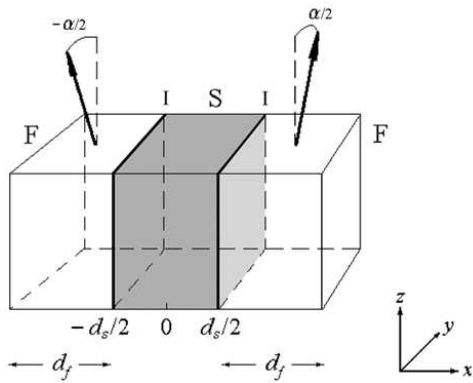}
    \caption{A schematic of an FSF junction with magnetization vectors lying in
                 $y$-$z$ planes, at angles $\pm\alpha/2$ with the $z$ axis.
                 The $x$-axis is perpendicular to the layer interfaces. The
                 strength of
                 insulating barriers (I) at the interfaces is measured by a
                 dimensionless
                 parameter $Z=2W_0/\hbar v_{\rm F}$.}
    \label{T1}
\end{center}
\end{figure}

We consider an FSF double junction consisting of a clean
superconducting (S) layer of thickness $d_s$, connected to clean
ferromagnetic layers (F) of thickness $d_f$ via thin insulating
interfaces (I), Fig.~\ref{T1}. To describe the ferromagnets we use
the Stoner model for inhomogeneous exchange field ${\bf h}({\bf r})$
lying in $y$-$z$ plane, parallel to the layers. The exchange field
has, in general, different orientations in the left and right
ferromagnet, which (without loss of generality) we take to be
symmetric with respect to the $z$-axis
\begin{equation}
    {\bf h}({\bf r}) = h_0 \left( 0, \pm\sin(\alpha/2),
    \cos(\alpha/2)\right),
\end{equation}
$ {\rm for}~|x|>d_s/2.$ The case $\alpha=0$ ($\alpha=\pi$) then
corresponds to the P (AP) alignment of magnetizations.

Quasiparticle propagation is described by the Bogoliubov--de
Gennes equation
\begin{equation}
\label{BdG0}
    \check{\cal{H}} \Psi({\bf r}) = E \Psi({\bf r})
\end{equation}
for the four-vectors $\Psi({\bf r}) \equiv \left(
  u_{\uparrow}({\bf r}) ~
  u_{\downarrow}({\bf r}) ~
  v_{\downarrow}({\bf r}) ~
  v_{\uparrow}({\bf r})
\right)^{\rm T}$, where the Hamiltonian can be written in the
compact form as
\begin{equation}
\label{Ham}
    \check{\cal{H}} = \widehat{\tau}_3 \otimes \left(\widehat{H}_0({\bf r}) - {\bf h}({\bf
    r})\cdot\widehat{\bm{\sigma}} \right) + \widehat{\Delta}({\bf
    r})\otimes\widehat{\bm{1}}.
\end{equation}
Here, $\widehat{\tau}_i$ and $\widehat{\sigma}_i$ are the Pauli matrices
in orbital and spin space, respectively, $\widehat{\bm{1}}$ is
$2\times2$ unity matrix,
\begin{equation}
    \widehat{H}_0({\bf r}) = \left(\begin{array}{cc}H_0({\bf r}) & 0 \\0 & H^\ast_0({\bf r})\end{array}\right),
\end{equation}
$H_{0}({\bf r})=-\hbar^{2}\nabla^{2}/2m+W({\bf r})+U({\bf
r})-\mu$ being the one-particle Hamiltonian, $U({\bf r})$ and $\mu$ are the
Hartree and the chemical potential, respectively, while $E$ is the energy with respect to $\mu$.
The interface potential is modeled symmetrically by $W({\bf r})=W_0\left[\delta(x+d_s/2)+\delta(x-d_s/2)\right]$,
where $\delta$ is the Dirac delta-function. The matrix $\widehat{\Delta}$ is given by
\begin{equation}
    \widehat{\Delta}({\bf r}) = \left(\begin{array}{cc}0 & \Delta({\bf r}) \\\Delta^\ast({\bf r}) & 0\end{array}\right),
\end{equation}
where $\Delta({\bf r})$ is the pair potential. We will not consider
the case of a ferromagnetic
superconductor,\cite{Eschrig,Eschrig05,Dong05,Marion} and henceforth we take
${\bf h}({\bf r})=0$ outside the ferromagnets and $\Delta({\bf
r})=0$ outside the superconductor. Since there is only one S layer
in the system, without loss of generality we can choose $\Delta$ to
be real. The electron effective mass $m$ is assumed to be the same
throughout the junction. For simplicity, the Fermi energy of the
superconductor and the mean Fermi energy of the ferromagnets are
assumed to be the same, $E_{\rm F}$. The model could be
straightforwardly generalized to the case of a Fermi
velocity mismatch between the layers (see,
e.g.~Ref.~\onlinecite{BozovicB}).

Wave-vector components have to be normalized by using the condition
for canonical transformation, which after integration over the
volume of the S layer reads
\begin{eqnarray}
\label{norm} \int {\rm d}^3{\bf r} \left[ |u_\sigma ({\bf r})|^2+
|v_{\bar{\sigma}} ({\bf r})|^2 \right] = 1.
\end{eqnarray}
Here, $\sigma$ denotes the spin orientation,
($\sigma=\uparrow,\downarrow$) and $\bar{\sigma}$ is opposite to
$\sigma$. As the exchange field removes spin degeneracy, a
spin-generalized gap equation has to be used.\cite{BozovicEPL} Using
the condition for the singlet pairing in the superconductor,
\begin{equation}
\label{sc1}
    \sum_{{\bf q}} \left[ u_{\uparrow}({\bf r})
    v^\ast_{\downarrow}({\bf r}) - u_{\downarrow}({\bf r})
    v^\ast_{\uparrow}({\bf r})\right] = 0,
\end{equation}
for $|x|\leq d_s/2$, the gap equation can be written in familiar
form
\begin{equation}
\label{sc2}
    \Delta({\bf r}) = V({\bf r}) \sum_{\bf q} \left[
    u_{\uparrow}({\bf r}) v^\ast_{\downarrow}({\bf
    r}) +
    u_{\downarrow}({\bf r}) v^\ast_{\uparrow}({\bf
    r})\right]\tanh\left(\frac{E}{k_{\rm B}T}\right).
\end{equation}
In Eqs.~(\ref{sc1}) and (\ref{sc2}) the summation is performed over wave vectors in the superconductor ${\bf q}$, taking into account the dispersion relation between ${\bf q}$ and $E$. Also, we assume for simplicity that the pairing interaction potential $V({\bf
r})=V=const.$ inside the superconductor.

The parallel component of the wave vector, ${\bf k}_{||,\sigma}$, is
conserved due to translational invariance of the junction in
directions perpendicular to the $x$-axis, but differs for the two spin
orientations. Consequently, the wave function can be written in the
form
\begin{equation}
    \Psi({\bf r}) =
    \left(\begin{array}{c}
      \widetilde{u}_{\uparrow}(x) e^{i{\bf k}_{||,\uparrow}\cdot{\bf r}} \\
      \widetilde{u}_{\downarrow}(x) e^{i{\bf k}_{||,\downarrow}\cdot{\bf r}}\\
      \widetilde{v}_{\downarrow}(x) e^{i{\bf k}_{||,\uparrow}\cdot{\bf r}}\\
      \widetilde{v}_{\uparrow}(x) e^{i{\bf k}_{||,\downarrow}\cdot{\bf r}}
    \end{array}\right).
\end{equation}
In the following, we will use the stepwise approximation for the
pair potential, $\Delta({\bf
r})=\Delta\Theta(d_s/2-x)\Theta(d_s/2+x)$, where $\Theta$ denotes
the Heaviside step function. In Ref.~\onlinecite{BozovicEPL} the averaged
self-consistent pair potential is calculated and can be used
instead of the bulk value. Fully self-consistent numerical
calculations have been performed recently for FSF\cite{Bagrets,Valls05} and
NS\cite{Bergmann} geometries (here N stands for a normal
non-magnetic metal).

The scattering problem for Eq.~(\ref{BdG0}) is solved assuming bulk
ferromagnetic layers, $d_f\to\infty$. Then, the four independent
solutions of Eq.~(\ref{BdG0}) correspond to the four types of
injection involving an electron or a hole incidence from either the
left or the right electrode.\cite{Furusaki Tsukada}

Let us first consider the electrons ($E>0$) with positive
$x$-components of the velocity. Solutions of Eq.~(\ref{BdG0}) are:

\noindent for the left F layer ($x<-d_s/2$)
\begin{eqnarray*}
    \widetilde{u}_{\uparrow}(x) &=&
        \cos(\alpha/2) e^{i k^+_\uparrow x}
        +\cos(\alpha/2) b_\uparrow e^{-i k^+_\uparrow x} \nonumber\\
        ~&~&-i\sin(\alpha/2) b_\downarrow e^{-i k^+_\downarrow x}, \\
    \widetilde{u}_{\downarrow}(x) &=&
        -i\sin(\alpha/2) b_\uparrow e^{-i k^+_\uparrow x}
        +\cos(\alpha/2) e^{i k^+_\downarrow x}\nonumber\\
        ~&~&
        +\cos(\alpha/2) b_\downarrow e^{-i k^+_\downarrow x}, \\
    \widetilde{v}_{\downarrow}(x) &=&
        -i\sin(\alpha/2) a_\downarrow e^{i k^-_\uparrow x}
        +\cos(\alpha/2) a_\uparrow e^{i k^-_\downarrow x}, \\
    \widetilde{v}_{\uparrow}(x) &=&
        \cos(\alpha/2) a_\downarrow e^{i k^-_\uparrow x}
        -i\sin(\alpha/2) a_\uparrow e^{i k^-_\downarrow x},
\label{TpsiL}
\end{eqnarray*}
for the S layer ($|x|<d_s/2$)
\begin{eqnarray*}
    \widetilde{u}_{\uparrow}(x) &=&
             \bar{u} c_1 e^{i q^+_\uparrow x}
        +\bar{u} c_2 e^{-i q^+_\uparrow x}
        +\bar{v} c_3 e^{i q^-_\uparrow x}
        +\bar{v} c_4 e^{-i q^-_\uparrow x}, \\
    \widetilde{u}_{\downarrow}(x) &=&
             \bar{u} c_5 e^{i q^+_\downarrow x}
        +\bar{u} c_6 e^{-i q^+_\downarrow x}
        +\bar{v} c_7 e^{i q^-_\downarrow x}
        +\bar{v} c_8 e^{-i q^-_\downarrow x}, \\
    \widetilde{v}_{\downarrow}(x) &=&
             \bar{v} c_1 e^{i q^+_\uparrow x}
        +\bar{v} c_2 e^{-i q^+_\uparrow x}
        +\bar{u} c_3 e^{i q^-_\uparrow x}
        +\bar{u} c_4 e^{-i q^-_\uparrow x}, \\
    \widetilde{v}_{\uparrow}(x) &=&
             \bar{v} c_5 e^{i q^+_\downarrow x}
        +\bar{v} c_6 e^{-i q^+_\downarrow x}
        +\bar{u} c_7 e^{i q^-_\downarrow x}
        +\bar{u} c_8 e^{-i q^-_\downarrow x},
\label{TpsiS}
\end{eqnarray*}
and for the right F layer ($x>d_s/2$)
\begin{eqnarray*}
    \widetilde{u}_{\uparrow}(x) &=&
        \cos(\alpha/2) c_\uparrow e^{i k^+_\uparrow x}
        +i\sin(\alpha/2) c_\downarrow e^{i k^+_\downarrow x}, \\
    \widetilde{u}_{\downarrow}(x) &=&
        i\sin(\alpha/2) c_\uparrow e^{i k^+_\uparrow x}
        +\cos(\alpha/2) c_\downarrow e^{i k^+_\downarrow x}, \\
    \widetilde{v}_{\downarrow}(x) &=&
        i\sin(\alpha/2) d_\downarrow e^{-i k^-_\uparrow x}
        +\cos(\alpha/2) d_\uparrow e^{-i k^-_\downarrow x}, \\
    \widetilde{v}_{\uparrow}(x) &=&
         \cos(\alpha/2) d_\downarrow e^{-i k^-_\uparrow x}
        +i\sin(\alpha/2) d_\uparrow e^{-i k^-_\downarrow x}.
\label{TpsiR}
\end{eqnarray*}
Here, $\bar{u}=\sqrt{(1+\Omega/E)/2}$ and
$\bar{v}=\sqrt{(1-\Omega/E)/2}$ are the usual BCS coherence
factors, and $\Omega=\sqrt{E^2-\Delta^2}$ is the quasiparticle
kinetic energy with respect to the Fermi level. Perpendicular
($x$-) components of the wave vectors in the F layers are
\begin{equation}
k^\pm_{\sigma}=\sqrt{\left(2m/\hbar ^2\right)\left(E_{\rm F}+\rho_{\sigma}h_0 \pm
E\right)-{\bf k}^2_{||,\sigma}},
\end{equation}
while in the S layer they are given by
\begin{equation}
q^\pm_{\sigma}=\sqrt{\left(2m/\hbar ^2\right)\left(E_{\rm F}\pm\Omega\right)-{\bf
k}^2_{||,\sigma}},
\end{equation}
where $\rho_\sigma=1~(-1)$ for $\sigma=\uparrow~(\downarrow)$. The
superscript ($+$ or $-$) corresponds to the sign of quasiparticle
energy.

At the layer interfaces the wavefunction $\Psi({\bf r})$ is
continuous, while its first derivative has a discontinuity
proportional to a dimensionless parameter $Z\equiv 2W_0/\hbar v_{\rm F}$
measuring the height of potential barriers at the interfaces,
\begin{eqnarray}
\label{bc1}
    \Psi({\bf r})|_{x=\pm (d_s/2\mp 0)} &=& \Psi({\bf r})|_{x=\pm(d_s/2\pm 0)},\\
    \frac{\partial\Psi({\bf r})}{\partial x}\Big|_{x=\pm (d_s/2\mp 0)} &=&
    \frac{\partial\Psi({\bf r})}{\partial x}\Big|_{x=\pm (d_s/2\pm 0)}\nonumber\\&~&-k_{\rm F} Z \Psi({\bf r})|_{x=\pm
    d_s/2}.
\label{bc4}
\end{eqnarray}
This yields a system of 16 linear equations in as many unknowns.
By solving it, we find the amplitudes $a_\sigma$, $b_\sigma$, $c_\sigma$, $d_\sigma$, $c_1, \ldots, c_8$.

The scattering of an electron coming from the left leads to four
possible processes: (1) local Andreev reflection (i.e., formation of
a Cooper pair by two electrons from the left F layer); (2) normal
reflection; (3) direct transmission to the right F layer; (4)
crossed, or nonlocal, Andreev reflection\cite{Russo,Beckmann} (i.e.,
formation of a Cooper pair by two electrons from the opposite F
layers). The respective probabilities are given by
\begin{eqnarray}
    \label{TA}
        A_\sigma&=&
        {\rm
        Re}\left(\frac{k^-_{\bar{\sigma}}}
        {k^+_\sigma}\right) |a_{\sigma}|^2  \nonumber\\
        ~&~& +\tan^2\left(\frac{\alpha}{4}\right) {\rm
        Re}\left(\frac{k^-_\sigma}
        {k^+_{\sigma}}\right) |a_{\bar{\sigma}}|^2, \\
    \label{TB}
        B_\sigma&=&
        |b_{\sigma}|^2 +
        \tan^2\left(\frac{\alpha}{4}\right) {\rm
        Re}\left(\frac{k^+_{\bar{\sigma}}}
        {k^+_{\sigma}}\right) |b_{\bar{\sigma}}|^2, \\
    \label{TC}
        C_\sigma&=&
        |c_{\sigma}|^2 +
        \tan^2\left(\frac{\alpha}{4}\right) {\rm
        Re}\left(\frac{k^+_{\bar{\sigma}}}
        {k^+_{\sigma}}\right) |c_{\bar{\sigma}}|^2, \\
    \label{TD}
        D_\sigma&=&
        {\rm
        Re}\left(\frac{k^-_{\bar{\sigma}}}
        {k^+_{\sigma}}\right) |d_{\sigma}|^2 \nonumber\\
        ~&~&+\tan^2\left(\frac{\alpha}{4}\right) {\rm
        Re}\left(\frac{k^-_\sigma}
        {k^+_{\sigma}}\right) |d_{\bar{\sigma}}|^2.
\end{eqnarray}
Conservation of probability current gives the usual normalization
condition
\begin{equation}
\label{ABCD} A_\sigma+B_\sigma+C_\sigma+D_\sigma=1.
\end{equation}
Solutions for the other three types of injection can be obtained
by symmetry arguments. In particular, the calculated probabilities
should be regarded as even functions of $E$.\cite{BozovicB}

It can be shown that $A_\sigma=D_\sigma=0$ whenever
\begin{equation}
\label{resonance}
d_s\left(q^+_{\sigma}-
q^-_{\sigma}\right)=2n\pi
\end{equation}
for $n=0,\pm 1,\pm 2,\ldots$, independently of $h_0$ and $Z$.
Therefore, both direct and crossed Andreev reflection vanish at
the energies of geometrical resonances in quasiparticle spectrum.
The absence of direct and crossed Andreev processes means that all
quasiparticles with energies satisfying Eq.~(\ref{resonance}) will
pass unaffected from one electrode to another, without creation or
annihilation of Cooper pairs. The effect is similar to the
over-the-barrier resonances in a simple problem of one-particle
scattering against a step-function potential,\cite{CohenT} the
superconducting gap playing the role of a finite-width
barrier.\cite{Tinkham}

Both the presence of insulating barriers and exchange interaction
reduce $A_\sigma$ and $C_\sigma$ and enhance $B_\sigma$ and
$D_\sigma$. Approaching the tunnel limit ($Z\to\infty$), the
spikes in $A_\sigma$, $C_\sigma$, and $D_\sigma$, as well as the
dips in $B_\sigma$, occur at the energies given by the
quantization conditions
\begin{equation}
\label{n} d_sq^+_{\sigma}=n_1\pi,~~~d_sq^-_{\sigma}=n_2\pi,
\end{equation}
which correspond to the bound-state energies of an isolated
superconducting film. In this case, these bound states are the
only conducting channels, both for supercurrent and
quasiparticle current.\cite{BozovicB,SPIE}

\section{Differential conductances}

\begin{figure}[h]
\begin{center}
    \includegraphics[width=7cm]{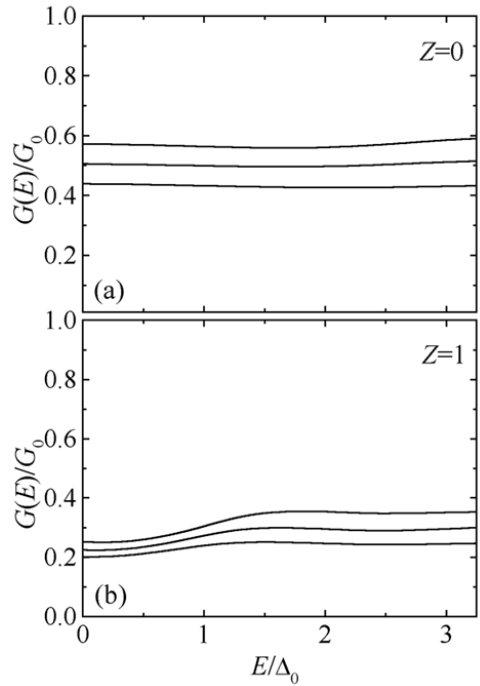}
    \caption{Differential conductance spectra of an FSF junction with (a) $Z=0$
                and (b) $Z=1$, for a thin S layer ($d_sk_{\rm F}=10^3$),
            $X=0.5$, and three relative orientations of magnetizations:
            $\alpha=0$, $\alpha=\pi/2$, and $\alpha=\pi$ (top to bottom).}
    \label{T2}
\end{center}
\end{figure}
\begin{figure}[h]
\begin{center}
    \includegraphics[width=7cm]{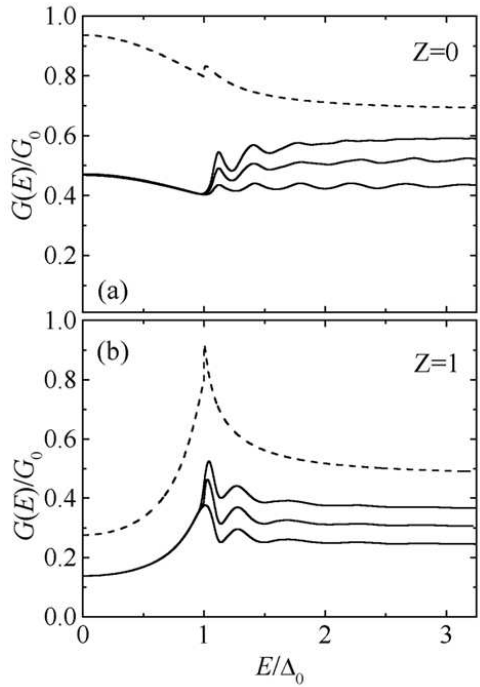}
    \caption{Differential conductance spectra of an FSF junction with (a) $Z=0$
                and (b) $Z=1$, for a thick S layer ($d_sk_{\rm F}=10^4$),
            $X=0.5$, and three relative orientations of magnetizations:
            $\alpha=0$, $\alpha=\pi/2$, and $\alpha=\pi$ (top to bottom).
            The spectra of the corresponding FIS
            junctions (generalized BTK results\cite{BozovicB}) are shown for
            comparison (dashed lines).}
    \label{T3}
\end{center}
\end{figure}

When voltage $V$ is applied to the junction symmetrically, the
charge current density can be written in the
form\cite{Yamashita67,Lambert,Zheng}
\begin{eqnarray}
J(V)&=&\frac{e}{(2\pi)^3 \hbar}\int\limits_{-\infty}^{\infty}{\rm
d}E \sum_{\sigma=\uparrow,\downarrow}P_\sigma \times\nonumber\\
~&~&\times\int {\rm d}^2{\bf
k}_{||,\sigma}
\left(1+A_\sigma-B_\sigma+C_\sigma-D_\sigma\right)\delta f({\bf
k},V)\nonumber\\
&=&\frac{e}{4\pi^3 \hbar}\int\limits_{-\infty}^{\infty}{\rm d}E
\sum_{\sigma=\uparrow,\downarrow}P_\sigma \times\nonumber\\
~&~&\times\int {\rm d}^2{\bf
k}_{||,\sigma} \left(A_\sigma+C_\sigma\right)\delta f({\bf k},V),
\end{eqnarray}
where $P_\sigma=(1+\rho_\sigma X)/2$ with $X=h_0/E_{\rm F}$, and
$\delta f({\bf k},V)$ is the asymmetric part of the nonequilibrium
distribution function of current carriers. In the last equality
the normalization condition, Eq.~(\ref{ABCD}), was taken into
account. Without solving the suitable transport equation we take
$\delta f({\bf k},V)=f_0(E-eV/2)-f_0(E+eV/2)$, where $f_0$ is the
Fermi-Dirac distribution function.\cite{Tinkham,BTK} In this
approach, the charge current per orbital transverse channel is
given by
\begin{equation}
\label{I} I(V)=\frac{1}{e}\int\limits_{-\infty}^{\infty}{\rm d}E
\left[f_0(E-eV/2)-f_0(E+eV/2)\right]G(E),
\end{equation}
where
\begin{equation}
\label{G} G(E) = G_0\sum_{\sigma=\uparrow,\downarrow}
    P_\sigma \int \frac{{\rm d}^2 {\bf k}_{||,\sigma}}{2\pi k_{\rm F}^2}
    \left(A_\sigma+C_\sigma\right)
\end{equation}
is differential charge conductance at zero temperature and
$G_0=e^2/\pi\hbar$ is the conductance quantum.

One can see that Eq.~(\ref{G}) is a simple generalization of the
Landauer formula,\cite{Landauer} with terms that take into account
transmission of the current through Cooper pairs and
quasiparticles. For $E<\Delta$, the subgap transmission of
quasiparticles (without conversion into Cooper pairs) suppresses
the Andreev reflection, while for $E>\Delta$ all the
probabilities oscillate with $E$ and $d_s$ due to the interference
of incoming and outgoing particles.

The influence of the exchange interaction and relative orientation
of magnetizations on the conductance spectra is illustrated for
$X=0.5$ and $Z=0$, for thin ($d_s k_{\rm F}=10^3$, Fig.~\ref{T2})
and thick ($d_s k_{\rm F}=10^4$, Fig.~\ref{T3}) S films. In all
the illustrations the bulk value of superconducting pair potential
is characterized by $\Delta_0/E_{\rm F}=10^{-3}$, which corresponds
to $\xi_0 k_{\rm F}=(2/\pi)\cdot 10^3$. For thin S layers, $d_s k_{\rm F} \lesssim
10^3$, the bulk value $\Delta_0$ is replaced by the averaged self-consistent pair potential, calculated in Ref.~\onlinecite{BozovicEPL}. The spin-polarized subgap transmission of
quasiparticles, and consequently a strong suppression of the Andreev
reflection, are significant in junctions with thin S films, whereas
the conductance oscillations above the gap become pronounced as
$d_s$ is increased. The magnetoresistance is apparent, as
conductance is greater for the P ($\alpha=0$) than for the AP
($\alpha=\pi$) alignment. The presence of non-collinear
magnetizations ($\alpha=\pi/2$) does not lead to any nonmonotonicity
in the conductance spectra with respect to collinear cases
($\alpha=0$ and $\alpha=\pi$) -- the values for $\alpha=\pi/2$
simply fall in between the P and AP curves. Moreover, it can be seen
that the conductances for $E>\Delta$ oscillate in phase for
$\alpha=0$, $\alpha=\pi/2$, and $\alpha=\pi$, Fig.~\ref{T3}(a). For
thin S films ($d_sk_{\rm F}\lesssim 10^3$) transmission of the spin
polarized current and suppression of Andreev reflection are still
dominant at energies below $\Delta$, Fig.~\ref{T2}(a). The
conductance spectra exhibit similar behavior for finite transparency
of the interfaces. This is illustrated for weak non-transparency
[$Z=1$, Figs.~\ref{T2}(b) and \ref{T3}(b)]. From both Figs.~\ref{T2}
and \ref{T3} it can be seen that conductances attain their
high-energy values, corresponding to conductances of an FNF double
junction, when $E$ is of the order of several $\Delta$.

\section{Magnetoresistance}

\begin{figure}[h]
\begin{center}
    \includegraphics[width=7cm]{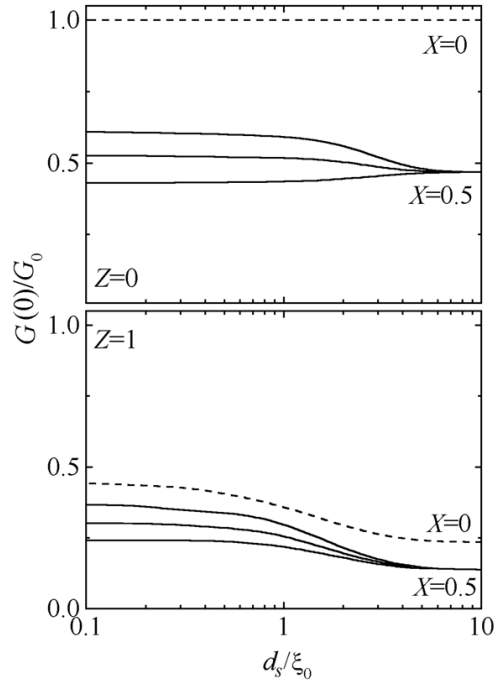}
    \caption{Zero-bias conductance of an FSF junction with $Z=0$ (top panel)
                and $Z=1$ (bottom panel) at zero temperature, as a function of
            the normalized S layer thickness $d_s/\xi_0$, for $X=0.5$,
            and three relative orientations of magnetizations:
            $\alpha=0$, $\alpha=\pi/2$, and $\alpha=\pi$ (top to bottom).
            The corresponding NSN values ($X=0$) are shown for comparison
            (dashed lines).}
    \label{T4}
\end{center}
\end{figure}

In clean FSF structures electrons propagating from F to S layer
are not necessarily converted into Cooper pairs. Moreover, part of
them has always a non-zero probability of being directly transmitted
from one F electrode to the other, even at the voltage below
$2\Delta/e$.\cite{BozovicB,Yamashita67} This probability increases
with decreasing $d_s$. As the applied voltage is increased, the
conductance spectrum starts to resemble the one of an FNF junction.
Fig.~\ref{T4} shows the zero-bias voltage $G(0)$ as a function of
 normalized thickness of the S layer, $d_s/\xi_0$, for three
relative orientations of magnetic moments: $\alpha=0$,
$\alpha=\pi/2$, and $\alpha=\pi$. It can be seen that mutual
differences between the corresponding values of $G(0)$ are larger
the thinner the S layers are. At thicknesses of the order of the
superconducting coherence length ($d_s/\xi_0\sim 1$) these
differences slowly disappear, and vanish completely at
$d_s/\xi_0\approx 5$. The increase of the normal reflection
probability with $d_s$ is more rapid than the increase of the
Andreev reflection probability if the insulating barriers were
present in the junction. Consequently, zero-bias conductances are
more sensitive to $d_s$ in the junctions without ($Z=1$) than in
those with the full transparency ($Z=0$), as can be seen in
Fig.~\ref{T4}.

\begin{figure}[h]
\begin{center}
    \includegraphics[width=7cm]{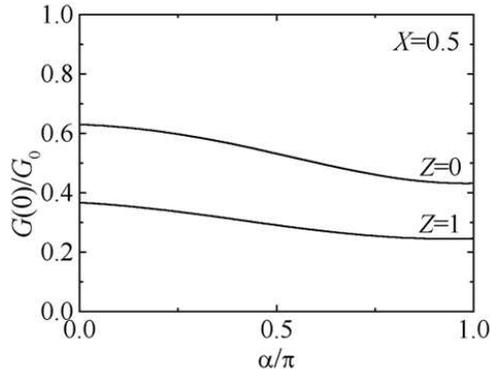}
    \caption{Zero-bias conductance of an FSF junction with $Z=0$
                and $Z=1$ at zero temperature, as a function of
            the relative orientation of magnetizations
            $\alpha$, for $X=0.5$ and thin S layer, $d_sk_{\rm F}=10^3$.}
    \label{T5}
\end{center}
\end{figure}

For hybrids with thin S layers ($d_s k_{\rm F}\sim 10^3$ or less) zero-bias
conductance decreases monotonously when $\alpha$ changes from $0$ to
$\pi$ (Fig.~\ref{T5}), both for $Z=0$ and $Z=1$. If the S layers are
thick ($d_sk_{\rm F}\sim 10^4$ or greater), $G(0)$ no longer depends on
$\alpha$: the Andreev reflection dominates at energies close to the
Fermi level, while the probability of direct transmission from any
of the F layers through the superconductor is negligible,
independently of mutual orientation of magnetizations. At energies
above the gap, dependence of $G(E)$ on $\alpha$ is also monotonous, with values close to those for the corresponding FNF junction.

\begin{figure}[h]
\begin{center}
    \includegraphics[width=7cm]{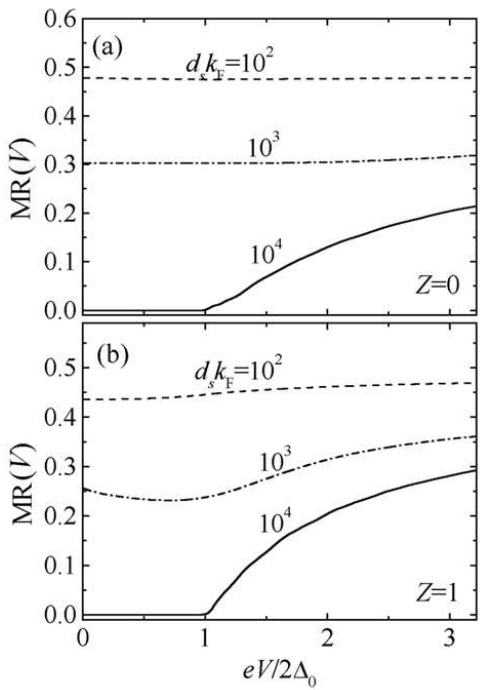}
    \caption{Magnetoresistance of an FSF junction with (a) $Z=0$
                and (b) $Z=1$ at zero temperature, as a function of the applied
            voltage, for $X=0.5$, and ultrathin, thin, and thick S film,
            $d_sk_{\rm F}=10^2,~10^3,~10^4$, respectively.}
    \label{T6}
\end{center}
\end{figure}

Magnetoresistance is defined as
\begin{equation}
    \label{MR}
    {\rm MR} \equiv \frac{R_{\rm AP}-R_{\rm P}}{R_{\rm P}},
\end{equation}
where $R_{\rm P(AP)}=V/I_{\rm P(AP)}$ is the junction resistance in
P (AP) alignment of magnetizations. In Fig.~\ref{T6}, MR is shown as
a function of bias applied symmetrically at the ends of an FSF
junction with $Z=0$ and $Z=1$, for $X=0.5$ and three values of
$d_sk_{\rm F}$: $10^2$, $10^3$, and $10^4$. The main characteristic
of junctions with thin S layers ($d_sk_{\rm F}\sim 10^2 - 10^3$)
is the dominance of direct transmission over Andreev reflection.
Also, the average number of quasiparticles converted to Cooper pairs
is small, independently of voltage, making the number of up spins that
cross from one F layer to the other much greater than the
corresponding down spins. This difference is greater in the P than
in the AP alignment of magnetizations, leading to a
magnetoresistance that gets more pronounced as the S layer becomes
thinner. Conversely, in junctions with a thick S layer ($d_sk_{\rm
F}\sim 10^4$), direct transmission probability is very low and hence
most of the electrons that enter the S layer at energies below
$\Delta$ are converted into Cooper pairs of net spin zero. Thus, for the electrons in one F layer the influence of the opposite F
layer becomes practically negligible, and therefore ${\rm MR}\simeq 0$ for
$eV/2\Delta_0\leq 1$. At higher voltages, $eV/2\Delta_0>1$, MR rises
practically monotonously due to a gradual increase of direct
transmission.

Therefore, magnetoresistance of clean FSF junctions exhibits a
strong dependence on the S layer thickness, even at a low bias. The
MR is always more pronounced if the exchange field is stronger.
However, it decreases towards zero with increasing $d_s$ due to the
dominance of supercurrent over the normal current.

\section{Spin-triplet correlations}

A peculiar property of proximity of singlet-pairing superconductors
and inhomogeneous ferromagnetic metals is inducement of triplet
correlations between electron-hole pairs of equal spin
orientations.\cite{BVE_rev} These correlations also exist in systems
with locally homogeneous exchange fields of non-collinear
configuration.\cite{VBE} Triplet correlations in F--S hybrids only
resemble those of magnetic superconductors: while in the latter the
Cooper pairs could be truly spin-triplet, in the former they remain
singlet despite the presence of both singlet and triplet components
of anomalous Green's function.

To study the triplet correlations in clean FSF junctions we solve
the Gor'kov equations generalized to take into account non-collinearity of
magnetizations in the F layers. In this case, the net Green's
function $\check{\mathcal{G}}({\bf r_1},{\bf r_2})$ is a $(2\times
2)\otimes(2\times 2)$ matrix in the Nambu space. The Gor'kov
equations can be compactly written as
\begin{equation}
\label{Trip Gorkov 4x4}
    \left(E\check{\mathbf{1}} - \check{\cal{H}}\right)
\check{\mathcal{G}}({\bf r_1},{\bf r_2}) = \delta({\bf
r_1}-{\bf r_2}) \check{\mathbf{1}},
\end{equation}
where $\check{\cal{H}}$ is given by Eq.~(\ref{Ham}),
while $\check{\mathbf{1}}$ is the unity matrix in the Nambu space. As the system of our
interest is part-by-part homogeneous, in each layer of the junction
the components of the matrix Green's function $\check{\mathcal{G}}$
are functions of the relative coordinate ${\bf r}\equiv {\bf
r_1}-{\bf r_2}$. It consists of four matrix blocks,
\begin{equation}
    \check{\mathcal{G}} = \left(\begin{array}{cc}\widehat{G}^{+} & \widehat{F} \\\widehat{F}^\ast & \widehat{G}^{-}\end{array}\right).
\end{equation}
Here, we are interested only in the anomalous block
\begin{equation}
\label{hatF}
    \widehat{F} = \left(\begin{array}{cc}F_{\uparrow\downarrow} &
F_{\uparrow\uparrow} \\F_{\downarrow\downarrow} &
F_{\downarrow\uparrow}\end{array}\right).
\end{equation}
The antisymmetric combination
$(F_{\uparrow\downarrow}-F_{\downarrow\uparrow})/2$ describes
singlet correlations, while the symmetric functions
$(F_{\uparrow\downarrow}+F_{\downarrow\uparrow})/2$,
$F_{\uparrow\uparrow}$, and $F_{\downarrow\downarrow}$ are related
to triplet correlations. If the exchange field ${\bf h}({\bf r})$
has the same direction in the left and the right F layer ($\alpha=0$), then
without loss of generality we can choose this to be direction of
the $z$ axis, see Fig.~\ref{T1}. The blocks $\widehat{H}_0({\bf r}) - {\bf h}({\bf
r})\cdot\widehat{\bm{\sigma}}$ in $\check{\cal H}$ are then diagonal,
and thus only the singlet correlations are present. The function
$F=F_{\uparrow\downarrow}=F_{\downarrow\uparrow}$ is then the
usual pair amplitude, while
$F_{\uparrow\uparrow}=F_{\downarrow\downarrow}=0$. In general, however, it can be the case that $0<\alpha<\pi$. Then, the
functions $F_{\uparrow\downarrow}$ and $F_{\downarrow\uparrow}$
are different, while $F_{\uparrow\uparrow}$ and
$F_{\downarrow\downarrow}$ are non-zero, leading to existence
of triplet correlations.

For ferromagnets, where
$\Delta =0$, Eq.~(\ref{Trip Gorkov 4x4}) reduces to
\begin{equation}
\left(E\widehat{\bf 1}-\widehat{H}_{+}({\bf r})\right) \widehat{F}({\bf r}) =0. \label{gorkovanomalous}
\end{equation}
General solution of this equation for the right F layer are
\begin{eqnarray}
    F_{\uparrow\downarrow} &=&
        \cos(\alpha/2) C_1 e^{i k^+_\uparrow x}
        +\cos(\alpha/2) C_2 e^{-i k^+_\uparrow x}\nonumber\\
        ~&~&+i\sin(\alpha/2) C_3 e^{i k^+_\downarrow x}
        +i\sin(\alpha/2) C_4 e^{-i k^+_\downarrow x},\\
    F_{\downarrow\downarrow} &=&
        i\sin(\alpha/2) C_1 e^{i k^+_\uparrow x}
        +i\sin(\alpha/2) C_2 e^{-i k^+_\uparrow x}\nonumber\\
        ~&~&+\cos(\alpha/2) C_3 e^{i k^+_\downarrow x}
        +\cos(\alpha/2) C_4 e^{-i k^+_\downarrow x},\\
    F_{\downarrow\uparrow} &=&
        i\sin(\alpha/2) C_5 e^{i k^-_\uparrow x}
        +i\sin(\alpha/2) C_6 e^{-i k^-_\uparrow x}\nonumber\\
        ~&~&+\cos(\alpha/2) C_7 e^{i k^-_\downarrow x}
        +\cos(\alpha/2) C_8 e^{-i k^-_\downarrow x},\\
    F_{\uparrow\uparrow} &=&
        \cos(\alpha/2) C_5 e^{i k^-_\uparrow x}
        +\cos(\alpha/2) C_6 e^{-i k^-_\uparrow x}\nonumber\\
        ~&~&+i\sin(\alpha/2) C_7 e^{i k^-_\downarrow x}
        +i\sin(\alpha/2) C_8 e^{-i k^-_\downarrow x},
\label{GorkovF}
\end{eqnarray}
modulo phase factor $\exp(i{\bf k}_{||,\sigma}\cdot {\bf r})$. Solutions in the left F layer are obtained if we
substitute $\alpha$ by $-\alpha$, and use a different set of
constants, $C_9, \ldots, C_{16}$. To find the complete set of
unknown constants $C_1, \ldots, C_{16}$ we have to use the
continuity of $F_{\sigma\sigma'}$ and its derivative at the S-F
interfaces. However, to apply the boundary conditions at $x=\pm
d_s/2$ we need to find solutions for $F_{\sigma\sigma'}$
inside the superconductor, where $G_{\sigma\sigma'}$ and
$F_{\sigma\sigma'}$ are coupled through the pair potential
$\Delta({\bf r})$. To remain within a tractable analytical
procedure, we use a simpler method which is sufficient for
qualitative discussion. First, we eliminate eight out of sixteen
constants using condition that holds at the outer boundaries,
\begin{equation}
\label{Fout}
    F_{\sigma\sigma'}\Big|_{x=\pm(d_s/2+d_f)} = 0.
\end{equation}
Then, we connect these functions with corresponding solutions in
the S layer, which for this purpose we treat as unknown
parameters, defined as
\begin{eqnarray}
\label{FA1}
    F_{\uparrow\downarrow}\Big|_{x=\pm d_s/2} = A_1,\\
    F_{\downarrow\uparrow}\Big|_{x=\pm d_s/2} = A_2,\\
    F_{\uparrow\uparrow}\Big|_{x=\pm d_s/2} = B_1,\\
    F_{\downarrow\downarrow}\Big|_{x=\pm d_s/2} = B_2.\label{FB2}
\end{eqnarray}
Note that the functions $F_{\sigma\sigma'}$ depend on $E$, ${\bf k_{||}}$, and $x$.
In order to describe the spatial variation of triplet components we introduce the pair amplitudes
\begin{equation}
\label{intF}
    f_{\sigma\sigma'}(x) = N(0)\int_0^\infty{\rm d}E\int\frac{{\rm d}^2{\bf
k}_{||,\sigma}}{2\pi k_{\rm F}^2} F_{\sigma\sigma'}(E,{\bf k_{||,\sigma}},x),
\end{equation}
where $N(0)$ is density of states at the Fermi level. Using these functions, we can further construct\cite{BVE_rev}
\begin{eqnarray}
\label{fff}
    f_0(x) &=& \frac{1}{2}\left[ f_{\uparrow\downarrow}(x) + f_{\downarrow\uparrow}(x) \right], \nonumber\\
    f_1(x) &=& f_{\uparrow\uparrow}(x) ~=~    f_{\downarrow\downarrow}(x),\\
    f_3(x) &=& \frac{1}{2}\left[ f_{\uparrow\downarrow}(x) - f_{\downarrow\uparrow}(x) \right] \nonumber.
\end{eqnarray}
The following symmetries have to hold: $f_0(-x)=f_0(x)$,
$f_1(-x)=-f_1(x)$, and $f_3(-x)=f_3(x)$. The function $f_3$
describes singlet, while $f_0$ and $f_1$ describe triplet
correlations in the system. The function $f_0$ corresponds to
electron-hole pairs of zero net spin orientation, while $f_1$
corresponds to pairs of net spin orientation equal to $\pm 1$. The
$f_3$ component exists even when the exchange field is absent. Both
$f_3$ and $f_0$ fall off rapidly in a ferromagnet: characteristic
decay lengths are of the order of $\xi_f=\sqrt{\hbar D_f/h_0}$ in
diffusive, and $\xi_f=\hbar v_{\rm F}/h_0$ in ballistic heterostructures,
where $D_f$ is a diffusion constant of a dirty ferromagnet. In
diffusive FSF junctions these components are of a short range, since
the exchange field tends to align spins, while $f_1$ is of a long
range and monotonically decaying.\cite{EfetovFSF} Here, we will
argue that in clean FSF hybrids $f_1$ can be of the same range as
$f_3$, and no long-range monotonically decaying triplet correlations
are generated.

\begin{figure}[h]
\begin{center}
    \includegraphics[width=7cm]{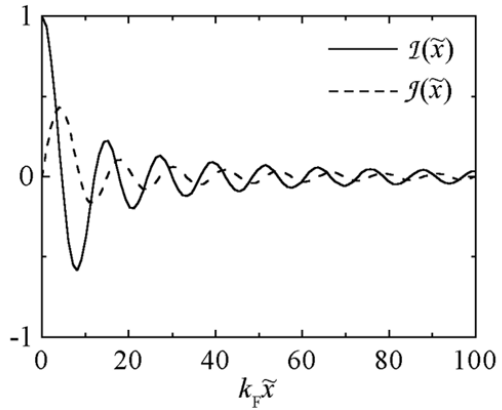}
    \caption{The functions ${\cal{I}}(\widetilde{x})$ (solid curve) and
            ${\cal{J}}(\widetilde{x})$ (dashed curve), under approximation
            $d_f\gg d_s$, shown for $\alpha=\pi/2$ and $X=0.5$.}
    \label{T11}
\end{center}
\end{figure}

If we assume that the F layers are bulk, $d_f\gg
d_s$, and focus on correlations in the vicinity of the S-F
interfaces, $|x|\gtrsim d_s/2$, then any plain wave of the form
$e^{\pm i k^\pm_\sigma d_f}$ will be rapidly oscillating with
respect to $e^{\pm i k^\pm_\sigma x}$. Therefore, the most important
contribution to the integral of
$F_{\sigma\sigma'}$ over $E$ in Eq.~(\ref{intF}) is from the energies close to $E=0$.
The energy dependence in the wavevectors is thus negligible, and we can write
$k^\pm_\sigma \simeq k^\pm_\sigma(E=0)\equiv k_\sigma$. Within this approximation the number of undetermined coefficients, defined by Eqs.~(\ref{FA1})--(\ref{FB2}), can be further reduced by applying the symmetries $f_{\uparrow\uparrow}(x) =
f_{\downarrow\downarrow}(x)$ and $f_{\uparrow\downarrow}(x) =
-f_{\downarrow\uparrow}(x)$ that now hold. Hence, $A_1=-A_2\equiv A$ and
$B_1=B_2\equiv B$. Performing the integration over ${\bf k}_{||,\sigma}$ in
Eq.~(\ref{intF}), we finally obtain the following set of expressions for the pair amplitudes of the right F layer
\begin{eqnarray}
    f_0(x) &=& 0 \label{f3clean0},\\
    f_1(x) &=& B{\cal{I}}(\widetilde{x}) + i \sin\alpha A{\cal{J}}(\widetilde{x}) \label{f3clean1},\\
    f_3(x) &=& A{\cal{I}}(\widetilde{x}) - i \sin\alpha B{\cal{J}}(\widetilde{x}).
    \label{f3clean3}
\end{eqnarray}
where $\widetilde{x}\equiv x-d_s/2$ is the relative distance from
the right S-F interface. In Eqs.~(\ref{f3clean0})--(\ref{f3clean3}) we have
introduced the auxiliary functions
\begin{eqnarray}
\label{I(x)}
    {\cal{I}}(\widetilde{x}) &=&
    \cos^2\left(\frac{\alpha}{2}\right) {\cal{Q}}_\uparrow(\widetilde{x})
    + \sin^2\left(\frac{\alpha}{2}\right) {\cal{Q}}_\downarrow(\widetilde{x}),\\
\label{J(x)}
    {\cal{J}}(\widetilde{x}) &=& {\cal{Q}}_\uparrow(\widetilde{x}) - {\cal{Q}}_\downarrow(\widetilde{x}),
\end{eqnarray}
where
\begin{equation}
    {\cal{Q}}_\sigma(\widetilde{x}) = \int \frac{{\rm d}^2 {\bf k}_{||,\sigma}}{2\pi k_{\rm F}^2} \frac{\sin\left[k_\sigma\left(d_f-\widetilde{x}\right)\right]}{\sin\left(k_\sigma d_f\right)}.
\end{equation}
The function ${\cal{I}}(\widetilde{x})$ is even and has a maximum at
the interfaces, ${\cal{I}}(0)=1$, while ${\cal{J}}(\widetilde{x})$
is odd, with ${\cal{J}}(0)=0$. The expressions for the pair amplitudes for the left F layer are straightforwardly obtained by substitutions
$\alpha\rightarrow-\alpha$ and $\widetilde{x}\rightarrow -\widetilde{x}$. For $\alpha=\pi/2$ the auxiliary functions, Eqs.~(\ref{I(x)}) and (\ref{J(x)}), then simplify to
\begin{eqnarray}
\label{I(x)s}
    {\cal{I}}(\widetilde{x}) &\simeq& 2 \left\lbrace \frac{\sin(\kappa \widetilde{x})}{\kappa \widetilde{x}} -
    \frac{1}{2}\left[\frac{\sin(\kappa\widetilde{x}/2)}{(\kappa\widetilde{x}/2)}\right]^2 \right\rbrace,\\
\label{J(x)s}
    {\cal{J}}(\widetilde{x}) &\simeq& \frac{\sin(\kappa\widetilde{x})}{(\kappa\widetilde{x})^2} -
    \frac{\cos(\kappa\widetilde{x})}{\kappa\widetilde{x}},
\end{eqnarray}
where $\kappa\equiv
k_{{\rm F}\uparrow}-k_{{\rm F}\downarrow}$ is the difference between
the Fermi wavenumbers for the two spin subbands,
\begin{equation}
\label{kappaXX}
    \kappa = k_{\rm F}\left[ (1+X)^{1/2}-(1-X)^{1/2} \right],
\end{equation}
and, as before, $X=h_0/E_{\rm F}$. The approximated functions ${\cal{I}}(\widetilde{x})$ and ${\cal{J}}(\widetilde{x})$, given by Eqs.~(\ref{I(x)s}) and (\ref{J(x)s}), are shown in
Fig.~\ref{T11} for $X=0.5$.

Under the assumptions made, the unknown parameters $A$ and $B$ in Eqs.~(\ref{f3clean1}) and (\ref{f3clean3}) will
depend on $X$ and $\alpha$. If $X=0$ and/or $\alpha=0,\pi$ then
$B=0$, since the triplet correlations are absent in these cases. From
Eqs.~(\ref{f3clean0})--(\ref{f3clean3}) we can draw the following
conclusions concerning triplet correlations in clean FSF
heterostructures. The function $f_0$ is approximately zero
inside the F layers. Since this function can only be generated
by the exchange field it also cannot exist in the S layer. The singlet
correlations, captured in $f_3$, consist of two terms. If $\alpha=0$ or $\pi$, $f_3$
reduces to $A{\cal{I}}(\widetilde{x})$. This term has a decay length
$2\pi/\kappa\sim\xi_f$. Thus, the range of singlet correlations
increases with decreasing strength of the exchange field. In special
case of an NSN junction ($X=0$) we obtain a well-known result that
singlet pair amplitude monotonically decays into the N layer.\cite{Valls04} For $X\neq 0$ and non-collinear
magnetizations two components of the triplet pair amplitude $f_1$ oscillate on the same scale and with the same decay length.

We conclude that in clean FSF junctions at zero temperature both
singlet and triplet pair correlations in ferromagnets are
oscillating and power-law decaying with the distance from S-F
interfaces. Physical intuition behind this result is that in such
systems there exists only one characteristic length determining
decay of correlations in the ferromagnets, which weakly depends on
excitation energy $E$. This stands in a contrast with findings for
diffusive FSF hybrids where two characteristic lengths are present:
$\xi_f$ and $\xi_\omega=\sqrt{D_f/2\hbar|\omega|}$, where
$\omega=(2n+1)\pi k_{\rm B}T$.
At temperatures just below the critical one for the
superconducting phase transition, $T<T_c$, the two lengths set up
two different scales, $\xi_\omega\sim \sqrt{\hbar D_f/2\pi k_B
T_c}\gg\xi_f$. Note that $\xi_\omega$ is also the length that
determines the decay of singlet component in nonmagnetic normal
metal.


\section{Conclusion}

We have studied the influence of misorientation of magnetizations on
the properties of ballistic transport in clean FSF trilayers. By
solving the Bogoliubov--de Gennes equation we have derived
generalized expressions for probabilities of processes that charge
carriers undergo. We use these probabilities to compute differential
conductances for arbitrary orientation of magnetic moments and
interface transparencies. Preferability of direct quasiparticle
transmission to the Andreev reflection in thin S layers and more
prominent resonant oscillations in thick S films are the main
consequences of quantum interference in clean
heterostructures. The subgap conductance is larger for P than for AP
alignment as a result of strong magnetoresistive effect in thin S
layers: when the S layer thickness is less or comparable to the
superconducting coherence length the direct transmission of spin
polarized quasiparticles across the superconductor becomes a
dominant transport mechanism. However, we show that no extraordinary
effects arise when the relative orientation of magnetizations is
between parallel and antiparallel -- the spectra for intermediary
angles simply fall in between those for P and AP alignment.

From the obtained results we can further draw some general
conclusions about the nature of ballistic transport in clean FSF
trilayers with inhomogeneous magnetizations. Firstly, the zero-bias
conductance depends on the relative orientation of magnetizations
only in heterostructures with thin superconducting layers, decreasing
monotonously when tuning from P to AP alignment. Secondly,
magnetoresistance displays qualitatively different voltage
dependance for thin and for thick superconducting layers due to
different quasiparticle spectra which tends to gapless or bulk,
respectively. Thirdly, magnetoresistance increases with the
thickness of the S layer and vanishes eventually. The effect is more
pronounced for strong exchange fields and transparent interfaces.

Non-collinearity of magnetizations leads to formation of spin-triplet
pair correlations in FSF structures. Unlike the diffusive case,
where the triplet correlations have a long-range and monotonically
decaying component, we have shown that in clean ferromagnet-superconductor
hybrids these correlations are oscillating and power-law
decaying with the distance from S-F interfaces, similarly to the
usual singlet correlations. This similarity in behavior of singlet
and triplet pair correlations induced in ferromagnets is the main
reason why the transport properties of clean FSF junctions have
monotonic dependence on the angle between magnetizations. These
findings suggest that any spectacular features of triplet
correlations are highly unlikely to occur in clean FSF structures.

\section*{ACKNOWLEDGMENT}

We thank Jerome Cayssol and Taro Yamashita for useful discussions. This work has been supported by the Serbian Ministry of Science, project No.~141014.


\end{document}